\documentclass[twocolumn]{aastex631}

\usepackage[utf8]{inputenc}
\usepackage{xcolor}

\begin{document}

\title{Discovery of Millisecond Pulsars toward the Galactic Bulge in an Image-based Survey with MeerKAT}

\author[0000-0002-9409-3214]{Rahul Sengar}
\affiliation{Max Planck Institute for Gravitational Physics (Albert Einstein Institute), D-30167 Hannover, Germany \\}
\affiliation{Leibniz Universit\"{a}t Hannover, D-30167 Hannover, Germany\\}
\affiliation{Center for Gravitation, Cosmology, and Astrophysics, Department of Physics and Astronomy, University of Wisconsin-Milwaukee, P.O. Box 413, Milwaukee, WI 53201, USA.\\}

\author[0000-0002-8935-9882]{Akash Anumarlapudi}
\affiliation{Department of Physics and Astronomy, University of North Carolina at Chapel Hill, 120 E. Cameron Ave, Chapel Hill, NC, 27599, USA\\}
\affiliation{Center for Gravitation, Cosmology, and Astrophysics, Department of Physics and Astronomy, University of Wisconsin-Milwaukee, P.O. Box 413, Milwaukee, WI 53201, USA.\\}

\author[0000-0001-6295-2881]{David L. Kaplan}
\affiliation{Center for Gravitation, Cosmology, and Astrophysics, Department of Physics and Astronomy, University of Wisconsin-Milwaukee, P.O. Box 413, Milwaukee, WI 53201, USA.\\}

\author[0009-0008-7604-003X]{Dale A. Frail}
\affiliation{National Radio Astronomy Observatory, P.O. Box O, Socorro, NM 87801, USA\\}

\author[0009-0006-5070-6329]{Scott D. Hyman}
\affiliation{Department of Engineering and Physics, Sweet Briar College, Sweet Briar, VA 24595, USA\\}

\author[0000-0003-3272-9237]{Emil Polisensky}
\affiliation{U.S. Naval Research Laboratory, 4555 Overlook Ave. SW, Washington, DC 20375, USA\\}

\

\begin{abstract}

We report on the follow-up observations of circularly polarized sources identified in the MeerKAT image-based survey of the Galactic bulge. Using the Parkes radio telescope, we observed sixteen circularly polarized sources with the UWL receiver and detected nine pulsars, six of which are new discoveries. All pulsars are fast rotators with spin periods under 100\,ms. Among the new  discoveries, five are millisecond pulsars (MSPs) and one has a spin period of 53.6\,ms. At least four new MSPs exhibit clear signs of binary motion in their discovery observations. The dispersion measures (DMs) of these pulsars fall between 18 and 330\,pc\,cm$^{-3}$, which are lower than expected for Galactic bulge members and indicate that these pulsars lie in the foreground along the line of sight rather than within the bulge itself. This is the first time such a large number of pulsars have been confirmed from candidates identified in an image-based survey. These discoveries underscore the exceptional efficacy of circular polarization selection in image-based pulsar surveys, and demonstrate the powerful synergy between high-sensitivity imaging and targeted time-domain follow-up using wide-band receivers, and strengthen prospects for future deep pulsation searches---e.g., with MeerKAT or the forthcoming SKA or DSA-2000---to uncover the true millisecond pulsar population in the Galactic bulge.

\end{abstract}


\section{Introduction} \label{sec:intro}

The Galactic center and its surrounding region, primarily the Galactic bulge, are thought to harbor a rich population of pulsars, including millisecond pulsars (MSPs) \citep[e.g.,][]{gronthier_18, berteaud_21}. The existence of such a pulsar population is not only of astrophysical interest but also carries significant implications for high-energy astrophysics. There is a long-standing debate concerning attempts to explain the reasons behind the Galactic-center GeV excess, which emerged when data taken by the Large Area Telescope (LAT) on board the \textit{Fermi} Gamma-ray Space Telescope revealed excess emission at energies of a few GeV near the Galactic center and extending several kiloparsecs into the Galactic bulge \citep{goodenough_09}.

Observationally, the Galactic bulge is commonly defined as the inner $\sim 10^\circ$ around the Galactic center ($|\ell| \lesssim 10^\circ$, $|b| \lesssim 10^\circ$), corresponding to a projected radius of $\sim 1.4$~kpc for $R_0 \simeq 8.1$--8.2~kpc. The GeV excess extends to at least $\sim 10^\circ$ and is frequently analyzed out to $\sim 20^\circ$ \citep{ajello_16}, implying substantial spatial overlap between the bulge and the inner regions of the excess. This also indicates that the GeV excess may extend beyond the bulge into the inner Galactic disk.

To explain this excess, two competing theories have been proposed. The first is that the excess of GeV emission is due to long-sought dark matter annihilation, since its spectrum and spatial distribution are broadly consistent with dark matter halo models \citep{abazajian_12}. The second theory suggests that the excess arises from a hidden population of MSPs in the Galactic bulge, whose gamma-ray spectra match the observed emission \citep{bartels_16}. Recently, statistical studies using \textit{Fermi}-LAT data have found that the gamma-ray emission near the Galactic-center region exhibits clumpiness that is more consistent with a population of faint, unresolved point sources rather than diffuse emission, thereby favoring the MSP population interpretation \citep{lee_16}. However, uncertainties in Galactic diffuse background modeling, due to interstellar emission remain too large to conclusively prove either the dark matter annihilation or the MSP population interpretation, thus leaving the issue unresolved \citep[e.g.,][]{calore_15, Cacciapaglia_20, dimauro_21}.

Expanding the pulsar census in the region near the Galactic center and Galactic bulge is therefore essential for clarifying the origin of the gamma-ray excess.  Population synthesis studies of MSPs suggest that over $10^4$ MSPs would be required to explain the gamma-ray excess \citep{2016ApJ...827..143C,gronthier_18}. However, despite extensive observational campaigns, searches for pulsars close to the Galactic center and bulge have yielded only a few detections \citep{deneva_09, johnston_06, Wongphechauxsorn_24}, and only one MSP has been reported within 1\degr\ of the Galactic Centre \citep{lower_24}. The scarcity of MSPs and the small number of slow pulsars can mainly be attributed to severe interstellar scattering and free--free absorption, which suppress the pulsed signal \citep[e.g.,][]{rajwade_17}, making them difficult to detect with standard time-domain searches. Moreover, the Galactic-center region is overwhelmingly crowded with bright, diffuse radio emission, primarily synchrotron radiation from cosmic-ray interactions in the Galactic magnetic field, which significantly raises the system temperature and thus reduces the sensitivity to faint point sources at low radio frequencies \citep[e.g.,][]{stappers_11}. Millimeter-wavelength and high frequency surveys designed to mitigate scattering effects have also failed to discover new pulsars. A key limitation to the non-discovery is that pulsar radio spectra are typically steep (magnetars are a notable exception) which reduces their detectability at high frequencies despite the reduced scattering. Therefore, the current pulsar census in the Galactic center and bulge remains incomplete and falls short of theoretical predictions, leaving the gamma-ray excess debate unresolved.

\begin{figure}
    \includegraphics[width=1.1\columnwidth]{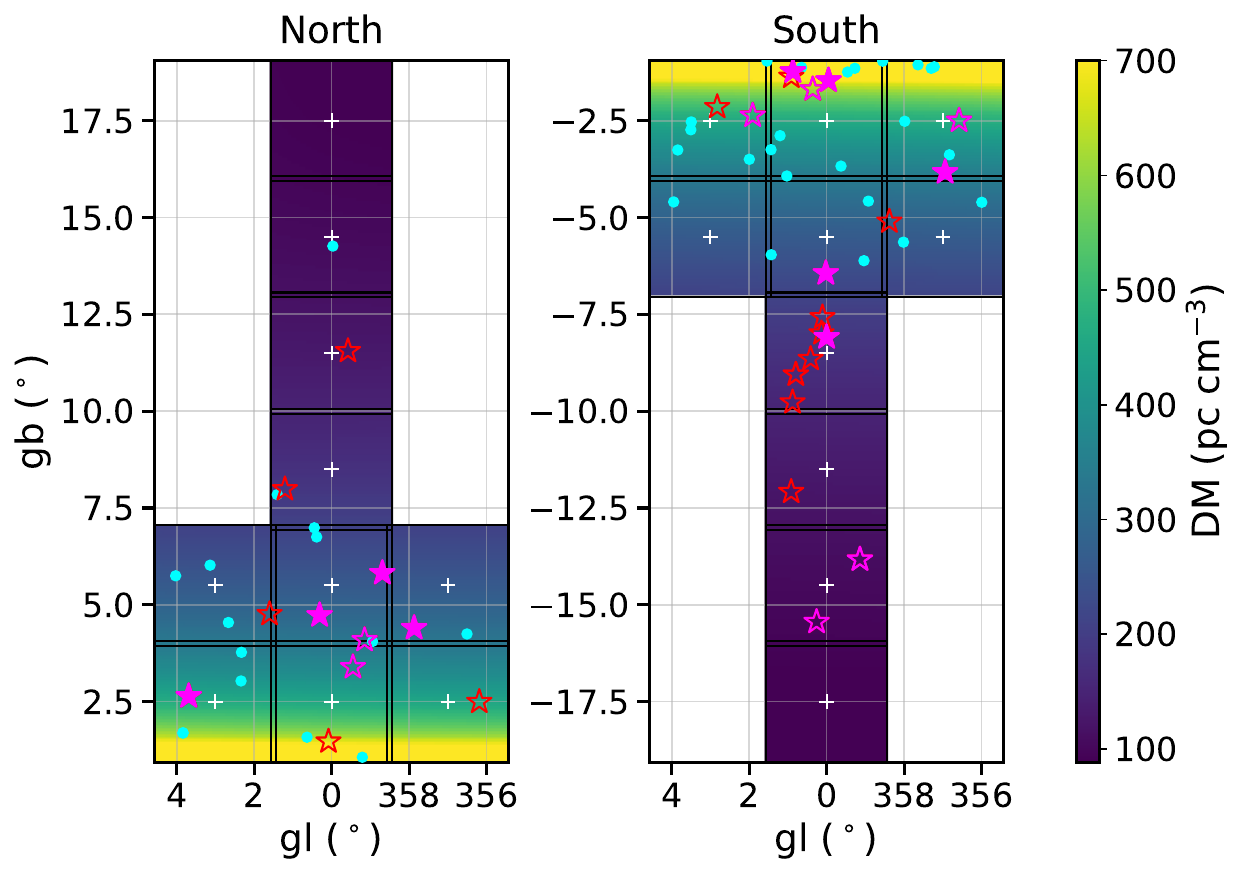}
     \caption{Distribution of the 20 MeerKAT mosaic pointings analyzed in radio-continuum images by \citet{frail_24}. The colored background shows the maximum line-of-sight DM from the YMW16 model, assuming a distance of 25 kpc. Brighter regions correspond to higher integrated electron column densities and thus higher DM. Gray crosses mark the centers of each pointing, each spanning approximately $3.125^\circ \times 3.125^\circ$. Previously known pulsars detected as polarized sources are shown as cyan points. Circularly polarized pulsar candidates are shown as stars: red stars indicate candidates not yet observed, magenta stars indicate candidates that have been observed, and filled magenta stars mark those confirmed as pulsars.}
    \label{fig:mosaic_pointings}
\end{figure}

\begin{deluxetable*}{cccccccccc}
\tablecaption{Properties of 16 circularly polarized sources we selected among 30 sources from \cite{frail_24} for the follow-up observations using the Parkes UWL receiver. For each source, the right ascension (RA), declination (DEC), Galactic longitude ($l$) and latitude ($b$), the measured Stokes I flux density at 1.4 GHz (I, in mJy), the spectral index ($\alpha$, derived from multi-frequency flux density fits), the fractional circular polarization (V/I), and the signal-to-noise ratio (S/N) of the detection in image plane are adopted from \citet{frail_24}. The total integration time (in seconds) and the observation date correspond to the Parkes follow-up observations. }
\label{tab:obs_sources}
\tablehead{
\colhead{RA} & \colhead{DEC} & \colhead{$l$} & \colhead{$b$} & 
\colhead{I} & \colhead{$\alpha$} & \colhead{V/I} & \colhead{S/N} & 
\colhead{Obs.\ Time} & \colhead{Obs. Date} \\
\colhead{(h:m:s)} & \colhead{(d:m:s)} & \colhead{($^\circ$)} & \colhead{($^\circ$)} & 
\colhead{(mJy)} & \colhead{} & \colhead{(\%)} & \colhead{} & 
\colhead{(s)} & \colhead{YR/MM/DD}
}

\startdata
17:20:18.40(1) & $-$26:52:07.8(2) &           $358.693$ & \phantom{0}$5.817$  & $0.20\pm0.01$ & $-1.87\pm0.31$ &         \phantom{-}$-27.8$  & \phantom{0}$7.9$ & 4320 & 2024/01/11 \\
17:23:29.79(1) & $-$28:20:36.1(1) &           $357.869$ & \phantom{0}$4.402$  & $0.59\pm0.01$ & $-1.52\pm0.11$ &          $-12.5$  & \phantom{0}$6.2$ & 3600 & 2025/01/28 \\
17:27:52.07(2) & $-$27:27:24.6(4) &           $359.147$ & \phantom{0}$4.093$  & $0.18\pm0.02$ & $-1.99\pm0.45$ & \phantom{-}\phantom{0}$26.4$ & \phantom{0}$6.2$ & 4320 & 2024/01/11 \\
17:28:20.26(1) & $-$26:08:16.6(2) & \phantom{00}$0.309$ & \phantom{0}$4.733$  & $0.22\pm0.01$ & $-2.90\pm0.26$ & \phantom{-}\phantom{0}$17.8$ & \phantom{0}$5.5$ & 4320 & 2024/01/11 \\
17:31:14.95(1) & $-$27:35:32.4(2) &           $359.446$ & \phantom{-}\phantom{0}$3.392$  & $0.30\pm0.01$ & $-2.18\pm0.22$ & \phantom{-}\phantom{0}$13.9$ & \phantom{0}$5.3$ & 3372 & 2025/03/28 \\
17:44:08.71(0) & $-$24:24:53.4(1) & \phantom{00}$3.686$ & \phantom{0}$2.640$  & $1.19\pm0.02$ & $-2.20\pm0.09$ & \phantom{-}\phantom{0}$19.9$ &         $18.8$    & 3960 & 2025/01/18 \\
17:47:18.05(0) & $-$33:09:17.4(1) &           $356.579$ &          $-2.497$           & $0.91\pm0.01$ & $-1.99\pm0.08$ & $-23.9$  &         $24.4$    & 3960 & 2025/01/18 \\
17:51:15.97(1) & $-$29:43:52.0(1) &           $359.953$ & $-1.464$            & $0.80\pm0.03$ & $-1.74\pm0.22$ & \phantom{-}\phantom{0}$8.3$ & \phantom{0}$5.6$ & 3504 & 2025/03/22 \\
17:52:29.06(1) & $-$28:49:05.6(1) & \phantom{00}$0.874$ & $-1.227$            & $1.07\pm0.02$ & $-0.86\pm0.12$ & \phantom{00}$9.0$ & \phantom{0}$7.1$ & 4320 & 2025/03/22 \\
17:53:06.42(2) & $-$29:30:00.8(3) & \phantom{00}$0.356$ & $-1.691$            & $0.26\pm0.02$ & $-1.51\pm0.51$ & \phantom{-}\phantom{0}$26.0$ & \phantom{0}$6.5$ & 3168 & 2025/01/19 \\
17:53:39.01(2) & $-$33:31:28.4(3) &           $356.939$ & $-3.826$            & $0.15\pm0.01$ & $-1.17\pm0.38$ & \phantom{-}\phantom{0}$23.7$ & \phantom{0}$5.4$ & 4320 & 2025/01/28 \\
17:59:20.04(3) & $-$28:29:56.6(3) & \phantom{00}$1.906$ & $-2.365$            & $0.13\pm0.01$ & $-2.24\pm0.57$ &          $-33.7$  & \phantom{0}$5.3$ & 3516 & 2025/02/26 \\
18:11:37.21(1) & $-$32:06:50.9(1) & \phantom{00}$0.022$ & $-6.436$            & $0.36\pm0.01$ & $-0.49\pm0.13$ & \phantom{-}\phantom{0}$17.4$ & \phantom{0}$9.3$ & 3444 & 2025/02/26 \\
18:18:31.06(1) & $-$32:54:18.2(2) &           $369.996$ & $-8.091$            & $0.20\pm0.01$ & $-1.18\pm0.25$ & \phantom{-}\phantom{0}$45.7$ &         $14.2$    & 3306 & 2025/03/28 \\
18:41:54.78(4) & $-$36:07:49.2(4) &           $359.140$ &          $-13.827$             & $0.14\pm0.01$ & $-1.58\pm0.55$ &    $-38.4$  & \phantom{0}$8.5$ & 5040 & 2025/03/22 \\
18:51:20.86(2) & $-$35:46:19.5(2) & \phantom{00}$0.255$ & $-15.442$       & $0.18\pm0.01$ & $-1.36\pm0.29$ & \phantom{-}\phantom{0}$20.0$ & \phantom{0}$5.3$ & 3048 & 2025/03/28 \\
\enddata

\tablecomments{The errors in RA and DEC are $1\sigma$ uncertainties.}

\end{deluxetable*}

In recent years, the development of high-sensitivity radio interferometers such as the Australian SKA Pathfinder \citep[ASKAP;][]{HOTAN_21}, Low Frequency Array \citep[LOFAR;][]{haarlem_13}, and MeerKAT \citep{frail_24} has enabled a complementary image-based pulsar search approach \citep[e.g.,][]{frail_24}. Instead of relying solely on periodicity detection, these searches exploit distinctive properties of pulsars as radio sources. Pulsars are generally compact, exhibit steep radio spectra, and more importantly they can display strong linear and/or circular polarization. Therefore, by identifying their compact, polarized, and steep-spectrum nature in interferometric images, one can obtain a set of sources exhibiting pulsar-like characteristics. This targeted selection provides an advantage over untargeted time-domain searches, which require hundreds to thousands of hours of sky coverage and, in dense Galactic regions such as the Galactic center or bulge, suffer from severe sensitivity losses. In time-domain searches, effects such as scattering smear out the pulses, and if the scattering timescales exceed the pulsar’s spin period, then the pulse essentially disappears into the background noise, making detection very difficult. However, imaging measures the time-averaged flux density of the source across the integration time. Thus, even though the pulsed signal is destroyed by scattering, the total flux is conserved. As a result, pulsars remain visible as compact and polarized sources, making image-based searches an effective complementary discovery method, as demonstrated by recent discoveries \citep[and references therein]{sengar_vast_25}.

\begin{figure}
    \includegraphics[width=0.99\columnwidth]{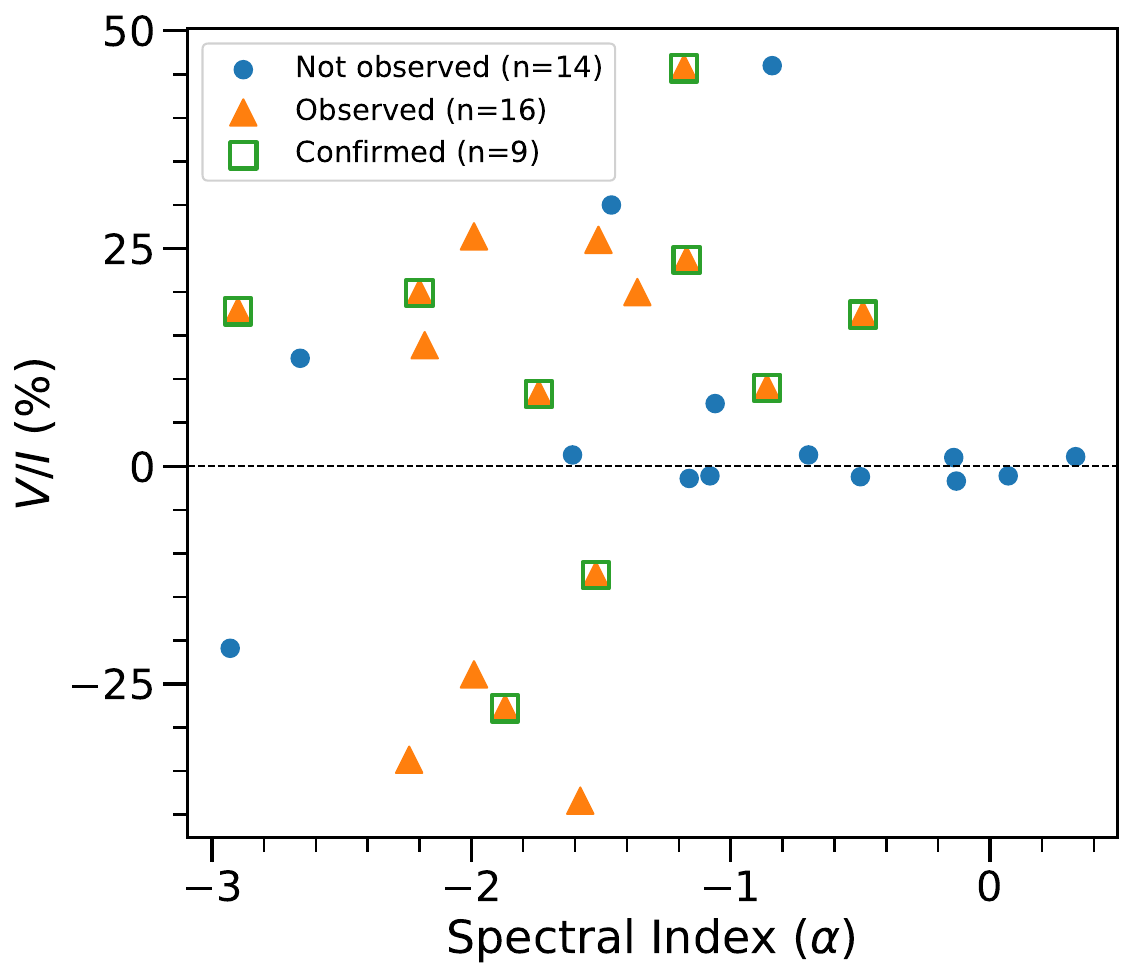}
     \caption{Spectral index, \(\alpha\) versus fractional circular polarization \(V/I\) for MeerKAT sources identified as circularly polarized in \citet{frail_24}. Orange triangles mark sources observed with the Parkes radio telescope; green open squares enclosing the triangles indicate sources confirmed in Parkes follow-up. Blue circles denote sources not yet observed. The horizontal dashed line marks \(V/I = 0\).}
    \label{fig:alpha_v_i}
\end{figure}

Building on this paradigm, \citet{cotton_25} presented a deep, full-Stokes MeerKAT L-band imaging survey of the Galactic bulge and central regions. The survey covered $\approx 173\,\mathrm{deg}^2$ at 856--1712\,MHz and reached root-mean-square noise levels of $\sim 12$--$17\,\mu\mathrm{Jy\,beam}^{-1}$.
The observations consisted of pointings mosaicked into 20 partially overlapped fields, each with a size of $3\fdg125 \times 3\fdg125$ (see Figure \ref{fig:mosaic_pointings}). Each mosaic was imaged in multiple subbands, enabling in-band spectral-index measurements and robust polarization calibration, and the resulting catalog includes full-Stokes polarization information. Using these survey products, \citet{frail_24} carried out an image-based search for pulsar-like sources that prioritized compact sources with pulsar-like polarization and spectral properties as candidates for time-domain follow-up. Applying these criteria to the mosaicked images yielded 30 circularly-polarized sources with properties resembling known pulsars. Complementary follow-up of these candidates has also yielded new MSPs \citetext{Maan et al., in prep.}\footnote{\url{http://www.ncra.tifr.res.in/~ymaan/scope.html}}. In this work, we report targeted pulsation searches of 16 candidates using the UWL receiver of the Parkes radio telescope.

The remainder of this paper is organized as follows. In Section \ref{sec:source_selection}, we summarize the candidate selection of the circularly polarized sources. Section \ref{sec:data_analysis} describes the Murriyang observations and data reduction for pulsation searches. In Section \ref{sec:discoveries}, we present new discoveries and rediscoveries of nine pulsars from FFT based periodicity and acceleration searches. Conclusion and discussion are presented in Section \ref{sec:discussion} where we discuss the potential reasons these pulsars were missed previously, and why the remaining candidates yielded no pulsar detections. We also highlight the importance and future prospects for image-based and circular polarization-focused searches in the era of the Square Kilometre Array.

\begin{figure*}
  \hspace*{0.025\textwidth}
  \includegraphics[width=1.95\columnwidth]{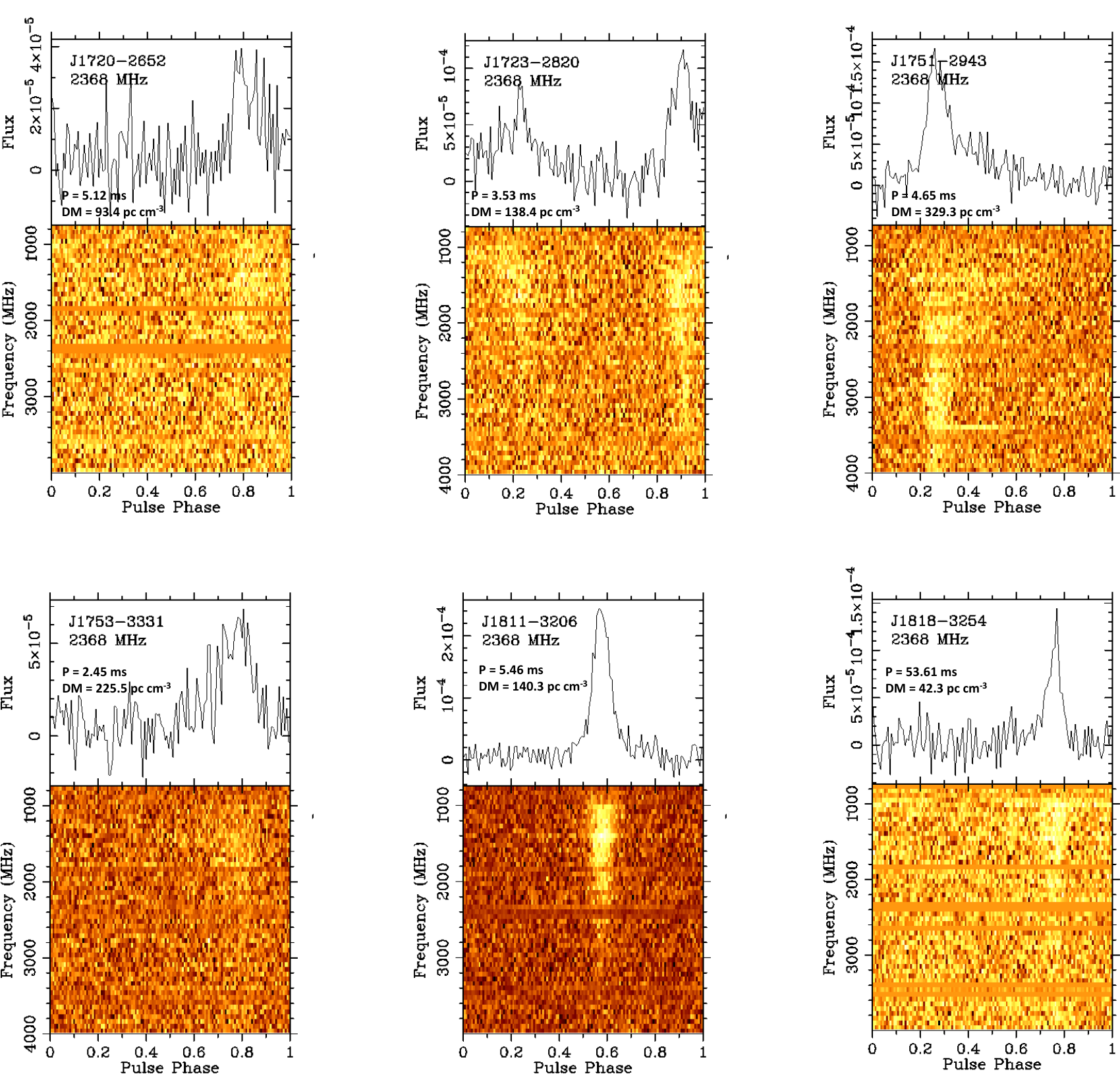}
  \caption{Discovery observations across the full UWL band (704--4032 MHz) for six newly discovered pulsars reported in this work. For each source, the upper sub-panel shows the phase-averaged total-intensity profile and the lower sub-panel shows the frequency-resolved intensity image. The horizontal axis is one full pulse phase (0--1). Profiles are in arbitrary flux units. All panels are generated with 128 phase bins and 52 frequency channels.}
  \label{fig:discovery_plots}
\end{figure*}

\section{Source selection } 
\label{sec:source_selection}

\citet{frail_24} reported 30 circularly polarized sources with measured \( |V|/I \) values that were not associated with multi-wavelength counterparts, and their other properties such as compactness, spectral index, and variability closely resembled those of the known pulsar population. These 30 pulsar candidates also included four low-band circularly polarized sources which were detected in the lower half of the MeerKAT L-band centered at 1.022 GHz. From these potential pulsar candidates, we selected 16 sources (see Table \ref{tab:obs_sources}) for targeted follow-up with the Parkes UWL receiver. Although all sources warrant confirmation attempts, observing constraints required us to prioritize a subset of these sources. However, our primary selection criterion was circular polarization which is now increasingly recognized as one of the main characteristic signatures of pulsars \citep[e.g.,][]{lenc_18, callingham_23}, while spectral signatures alone have a high false-positive rate \citep[e.g.,][]{kaplan_2000, froney_2000, maan_18}.

In particular, polarimetric studies of large samples of known pulsars have shown median fractional circular polarization values of \( |V|/I \sim 8\text{--}10\% \), with a tail extending to 20--30\% and, in some cases, even higher \citep[see Figure ~11 in][]{anumarlapudi_23}. By contrast, other common continuum sources, such as active galactic nuclei (AGN) cores and star-forming regions, typically show negligible fractional circular polarization (\(\lesssim 0.5\%\)) \citep[e.g.,][]{rayner_2000, homan_06}. Thermal or synchrotron emission from star-forming regions and H~II regions is expected to be unpolarized in circular polarization, with upper limits at the \(\lesssim 0.5\%\) level in galaxy surveys \citep[e.g.,][]{beck_15, irwin_18}. In practice, this approach reduces the number of plausible pulsar candidates from several thousand detections in the MeerKAT mosaic pointings to only a few dozen circularly polarized sources with properties resembling those of pulsars.

From 30 candidates, we identified 21 sources with $|V|/I > 5\%$, for which the mean fractional circular polarization is $\langle |V|/I\rangle = 23\%$. As a secondary discriminator, we also considered the radio spectral index. This subset of 21 sources has a mean spectral index of $\langle \alpha \rangle = -1.7 \pm 0.3$, consistent with the steep spectra typically observed for pulsars and consistent with average values reported in the ATNF pulsar catalogue across different spin-period ranges. For reference, the mean ATNF spectral indices are of order $\alpha \approx -1.9$ for MSPs ($P<30$ ms), $\alpha \approx -1.6$ for $30<P<100$ ms pulsars, and $\alpha \approx -1.8$ for slower pulsars ($P>100$ ms). These 21 sources therefore formed our highest-priority set for confirmation. By contrast, the remaining candidates have substantially lower fractional circular polarization and markedly flatter spectra, with $\langle |V|/I\rangle = 1.2\%$ and $\langle \alpha \rangle = -0.6$, which are more characteristic of common contaminant populations (e.g., compact AGN cores) than of pulsars. Accordingly, by combining (i) relatively high fractional circular polarization and (ii) steep, pulsar-like spectral indices, we selected 21 sources that satisfy both criteria (see Figure \ref{fig:alpha_v_i}). However, owing to limited available telescope time, we ultimately conducted targeted Parkes follow-up observations for 16 of them.

\section{Murriyang observations and data analysis} 
\label{sec:data_analysis}

To confirm whether any of the circularly polarized candidates were pulsars, we conducted follow-up observations with the Parkes 64-m radio telescope using the Ultra-Wideband Low (UWL) receiver \citep{hobbs_20} under project ID P1336. The UWL provides continuous frequency coverage from 704\,MHz to 4032\,MHz, allowing simultaneous searches across both low and high frequency ranges. This wide bandwidth is particularly advantageous for detecting steep-spectrum and faint pulsars at lower frequencies (e.g., 700--1500\,MHz), as well as mitigating the effects of interstellar scattering that can render pulsars undetectable in conventional L-band surveys.

\begin{deluxetable*}{lccccccccc}
\tablecaption{Properties of pulsars discovered and rediscovered in this work toward the Galactic bulge. Period, DM and folded S/N of the pulsars are those obtained from periodicity searches. Distances are derived from DM using the NE2001 and YMW16 free-electron density models.}
\label{tab:msps}
\tablewidth{0pt}
\tablehead{
\colhead{JName} & \colhead{RA (J2000)} & \colhead{DEC (J2000)} &
\colhead{$\ell$} & \colhead{$b$} &
\colhead{Period} & \colhead{DM} &
\colhead{$d_{\rm NE2001}$} & \colhead{$d_{\rm YMW16}$} &
\colhead{Folded S/N} \\
\colhead{} & \colhead{(hh:mm:ss)} & \colhead{(dd:mm:ss)} &
\colhead{(deg)} & \colhead{(deg)} &
\colhead{(ms)} & \colhead{(pc cm$^{-3}$)} &
\colhead{(kpc)} & \colhead{(kpc)} & \colhead{}
}
\startdata
J1720$-$2652$^{\bigstar}$ & 17:20:18.38(1) & $-$26:52:07.4(2) & 358.693 & \phantom{$-$}5.817 & \phantom{0}5.12759 & \phantom{0}93.4 & 2.0 & 2.9 & 12.3 \\
J1723$-$2820$^{\bigstar}$ & 17:23:29.79(1) & $-$28:20:36.1(1) & 357.870 & \phantom{$-$}4.402 & \phantom{0}3.53346 & 138.4 & 2.7 & 3.8 & 19.5 \\
J1744$-$2424$^{a}$ & 17:44:08.71(0) & $-$24:24:53.4(1) & \phantom{00}3.686 & \phantom{$-$}2.640 & 11.75691 & 196.8 & 3.5 & 4.4 & 26.0 \\
J1751$-$2943$^{\bigstar}$ & 17:51:15.97(1) & $-$29:43:52.0(1) & 359.954 & $-$1.464 & \phantom{0}4.65359 & 329.3 & 4.8 & 4.9 & 28.3 \\
J1752$-$2849$^{b}$ & 17:52:29.06(1) & $-$28:49:05.6(1) & \phantom{00}0.875 & $-$1.227 & 85.85005 & \phantom{0}18.4 & 0.6 & 0.7 & 13.0 \\
J1753$-$3331$^{\bigstar}$ & 17:53:39.01(2) & $-$33:31:28.4(3) & 356.939 & $-$3.827 & \phantom{0}2.45464 & 225.5 & 4.7 & 8.6 & 14.5 \\
J1811$-$3206 & 18:11:37.21(1) & $-$32:06:50.9(1) & \phantom{00}0.023 & $-$6.436 & \phantom{0}5.46334 & 140.3 & 3.5 & 6.1 & 48.0 \\
J1818$-$3254 & 18:18:31.06(1) & $-$32:54:18.2(2) & 359.996 & $-$8.091 & 53.61610 & \phantom{0}42.3 & 1.1 & 1.1 & 16.3 \\
\hline
J1728$-$2608$^{c}$ & 17:28:20.24(1) & $-$26:08:16.2(1) & \phantom{00}0.309 & \phantom{$-$}4.734 & \phantom{0}23.62019 & 138.4 & 4.1 & 2.8 & 11.5 \\
\enddata
\tablecomments{Distances are model-dependent estimates. $^{\bigstar}$ Pulsar most likely in a binary system. $^{a}$ Unpublished pulsar in a binary system initially identified as PSR J1743--2427 due to positional uncertainty and unavailability of phase connected timing solution \citep{bala_22_phd}. $^{b}$ Redetection of PSR J1753--2851 which was previously reported as PSR J1753--28 \citep{cameron_20}.$^{c}$ Rediscovery, originally found by Maan et al. (in prep.)}

\end{deluxetable*}

We observed each source in pulsar search mode (Stokes~I) for between 50 and 80 minutes. Data were recorded with a time resolution of 64\,\textmu s, 2-bit sampling, and a channel width of 500\,kHz, yielding 6656 frequency channels across the band. To further improve sensitivity across the wide UWL band, we employed a sub-banded search strategy: each full-band observation was divided into 12 subbands, comprising eight with 418\,MHz bandwidth and four with 832\,MHz bandwidth. This approach preserves frequency-dependent signal properties and helps to recover pulsars that may only be detectable in restricted frequency ranges. DMs were searched up to 1500\,pc\,cm$^{-3}$, corresponding to twice the maximum line-of-sight values predicted by the NE2001 \citep{ne2001} and YMW16 \citep{ymw16} electron density models.

The periodicity and acceleration searches followed the methodology described in \citet{sengar_23, sengar_htru_25}, which has proven highly effective in increasing the survey yield of multiple large scale surveys via their reprocessings. In brief, we used the GPU-accelerated search code \texttt{PEASOUP}\footnote{\url{https://github.com/ewanbarr/peasoup}}, which employs a time-domain resampling acceleration search technique under the constant-acceleration approximation where the line-of-sight (los) acceleration can be treated as approximately constant across the integration. This technique is near-optimal when the integration time, $t_{\rm int}$, spans a small fraction of the orbital period, $P_{\rm orb}$, typically $t_{\rm int} \lesssim 0.1\,P_{\rm orb}$, so that the los acceleration can be treated as approximately constant during the observation. For longer integrations ($t_{\rm int} \gtrsim 0.1\,P_{\rm orb}$), the acceleration can vary significantly, and can violate this assumption and causing signal smearing due to higher-order (jerk) effects. With our typical $t_{\rm int}\approx 60$\,min, the sensitivity begins to decline for binaries with $P_{\rm orb} \lesssim 10$\,hr, with increasingly severe degradation as $t_{\rm int}/P_{\rm orb}$ increases. To account for binary motion within the constant-acceleration framework, we searched acceleration trials up to $\pm 25$\,m\,s$^{-2}$.

To quantify the range of systems covered by this range, we consider circular binaries in the edge-on case ($\sin i = 1$) and require both $P_{\rm orb} \ge 10$\,hr (to satisfy the 10\% rule for a 60\,min integration) and $|a_{\rm los}| \le 25$\,m\,s$^{-2}$. Under these assumptions, the corresponding companion masses are $\lesssim 1.0\,M_\odot$ for a $1.4\,M_\odot$ pulsar. For comparison, a double neutron-star system with $m_c = 1.4\,M_\odot$ and $P_{\rm orb} = 10$\,hr has a maximum los acceleration of $\sim 35$\,m\,s$^{-2}$, which exceeds the acceleration range searched here. Nevertheless, the mismatch is modest, and our acceleration coverage remains adequate for binaries with moderate orbital periods and accelerations. Our search is not intended to be complete for the most strongly accelerated compact MSP binaries in 50--80\,min integrations, for which jerk searches or more comprehensive orbital-search methods provide higher sensitivity. To improve sensitivity to narrow duty-cycle pulsars, harmonic summing up to the 32nd harmonic was applied. Candidates down to a spectral signal-to-noise threshold of 5.5 were retained, typically producing $\sim$15,000 candidates per observation.  

Since it is impractical to fold all candidates, we employed a selection strategy that did not rely solely on ranking by spectral S/N. Instead, heuristic criteria based on candidate spin period, number of harmonics, and spectral S/N were applied to select candidates from both the strongest detections and the FFT noise floor. This approach reduced the candidate set to a manageable number for folding. Folding was performed with the \texttt{dspsr} package, with radio frequency interference excision applied using \texttt{clfd}. Diagnostic plots were generated with \texttt{pdmp}, and final candidate evaluation followed the procedures described in \citet{sengar_23}.

In addition to the FFT-based acceleration searches, we also conducted periodicity searches using the Fast Folding Algorithm \citep[FFA;][]{staelin_69} implemented in the \texttt{RIPTIDE}\footnote{\url{https://github.com/v-morello/riptide}} package \citep{morello_20}, as well as single-pulse searches to identify rotating radio transients (RRATs). For the FFA searches, dedispersed time series were generated with \texttt{PRESTO}'s\footnote{\url{https://github.com/scottransom/presto}} \texttt{prepsubband} routine up to a maximum DM of 1500 $ \rm pc \, cm^{-3}$ for each of the sub-banded segments. These time series were then searched for periodicities using \texttt{RIPTIDE}. Since FFT-based searches are generally more sensitive to MSPs, the FFA search range was restricted to spin periods between 100 ms and 30 s, targeting longer-period pulsars that may be missed in the Fourier domain. Candidate signals with FFA S/N $>$~9 were subsequently folded and evaluated using the same diagnostic criteria applied in the FFT searches. Each candidate was visually inspected, but no convincing candidate was detected. For the single-pulse searches, we again used the same dedispersed time series data generated by \texttt{prepsubband}. The \texttt{single\_pulse\_search} tool from the \texttt{PRESTO} package was employed to identify bright, individual bursts over a wide range of DMs. However, this search also did not yield any convincing candidate. Note that both the FFA and single pulse searches were conducted for isolated pulsars and time-series were not resampled for acceleration trials. While slow pulsars in binaries are less likely, they are not impossible. Therefore, acceleration searches with the FFA are subject to future work. 

\section{Discoveries}
\label{sec:discoveries}

From FFT-based periodicity searches, we detected periodic emission from nine sources out of the sixteen circularly polarized pulsar candidates observed with Murriyang (Parkes) using the UWL receiver, yielding a discovery success rate of $\sim 60\%$ for this sample. The spin periods of these pulsars range from $2.46$\,ms to $85.85$\,ms and their DMs are in the range $18.4 \lesssim {\rm DM} \lesssim 329.4\,{\rm pc\,cm^{-3}}$. Six of the nine pulsars are new discoveries, among which five have $P < 6$\,ms and are therefore classified as MSPs, while the remaining pulsar is also relatively fast with a spin period of $P \approx 53.6$\,ms. Of the remaining three pulsars, one is an independent discovery and the other two are redetections. Among the six new pulsars, we identified four MSPs showing orbital modulation, strongly indicating that they are likely to reside in binary systems. None of these MSPs show high acceleration in their detection data, but only low values ranging from $0.07 \,\mathrm{m\,s^{-2}} < a < 0.52 \,\mathrm{m\,s^{-2}}$. Using Kepler's third law, we explored the corresponding parameter space of orbital period and companion mass. The observed accelerations are consistent with binary systems hosting low-mass companions, either in the form of ``spider'' MSPs or typical He--white dwarf binaries with $M_c \lesssim 0.2\,M_\odot$. Assuming companion masses in the range $0.05\!-\!0.20\,M_\odot$, the implied orbital periods lie between $\sim 1$ and $5$\,days for $M_c \approx 0.05\,M_\odot$, and between $\sim 3$ and $12$\,days for $M_c \approx 0.20\,M_\odot$ which suggests that these binaries likely have orbital periods of several days, rather than in compact orbital systems. We do not yet have timing solutions for these pulsars. Continued timing observations will be essential to determine their precise orbital parameters and confirm the nature of their companions. However, since these pulsars were identified in imaging with sub-arcsecond positional accuracy, phase-connected timing solutions should be achievable within only a few months of follow-up.

The basic properties of these pulsars are summarized in Table~\ref{tab:msps} where we provide their imaging coordinates, spin period, DM, DM-derived distances from the NE2001 and YMW16 electron density models, and the folded S/N. The discovery plots of these six new pulsars including their pulse profiles are shown in Figure \ref{fig:discovery_plots}. These plots illustrate these pulsars exhibit a broad range of spectral characteristics, from relatively low to high S/N. In particular, PSRs~J1720$-$2652 and J1753$-$3331 are the faintest detections, with S/N values of 12 and 14, respectively, across their usable $1200$\,MHz bandwidth (1000--2200\,MHz). Both sources appear to be highly scattered at L-band frequencies. If they had been observed with a $\sim$500 MHz bandwidth typical of standard L-band receivers (e.g., 1000--1500\,MHz), their expected S/N would have been only $\sim$8--10, making them significantly harder to detect. PSR~J1723$-$2820 displays a double-peaked profile with components separated by approximately $250^\circ$ in pulse phase. PSR~J1751$-$2943 appears to be the most strongly scattered among the new discoveries, yet remains detectable across nearly the entire UWL band (1000--4032\,MHz). PSR~J1811$-$3206 is the brightest source in the sample, with an S/N of 48 and a notably steep spectrum. Its duty cycle of $\sim 8\%$ combined with its high S/N and narrow profile make it a strong candidate for inclusion in pulsar timing arrays (PTAs). PSR~J1818$-$3254 also has a steep spectrum and the narrowest profile in the sample. Its spin period of $53.6$\,ms is consistent with either a partially recycled pulsar (typically associated with double neutron star systems) or a young pulsar with a relatively large spin-down rate. Although no significant acceleration was detected in its discovery observation, suggesting that the pulsar is either isolated or in a wide-orbit binary, long-term timing will be required to measure $\dot{P}$ and thereby distinguish between these possibilities. We also note that a young-pulsar interpretation could imply an association with a supernova remnant or pulsar wind nebula; dedicated timing will be useful to assess this scenario.

During our 2024 campaign, we observed the MeerKAT source 172820.26$-$260816.6, however, we did not thoroughly inspect the faint candidates at that time. However, this source was later observed by another group (Maan et al., in prep.) who reported it as a 23.6 ms pulsar. Upon revisiting our results, we identified the same source with a faint detection due to RFI in that particular observation. We have included this source in Table \ref{tab:msps}, but we emphasize that this is a rediscovery rather than a new discovery.

The two other pulsars we detected are previously known sources, namely PSRs J1744$-$2424 and J1752$-$2849, with spin periods of 11.75\,ms and 85.8\,ms, and DMs of 196.8 and 18.4\,$\rm pc\,cm^{-3}$, respectively. PSR J1744$-$2424 is an unpublished pulsar initially identified in the processing of the HTRU survey and later redetected in a reprocessing of the HTRU-S LowLat survey \citep{bala_22_phd}. In the absence of a phase-connected timing solution, only preliminary binary parameters were available. This pulsar is in a 70.7-day orbit and has a minimum companion mass of $M_c \approx 0.04\,M_\odot$, making it an interesting binary system. Our detection suggests that this pulsar is associated with the source J174408.71$-$242453.4. 

In the observation of the source J175229.06$-$284905.6, we detected a pulsar with a period of 85.8\,ms. This source had previously been reported as J1753$-$28, though without an established timing solution \citep{cameron_20}. Unpublished timing data (A. Cameron, {\it priv. comm.}) now places this pulsar at the same position as one of the polarized sources, J175306.50$-$285126.2 from \citet{frail_24}. As this source was not included in our sample, this implies that our detection of the 85.8\,ms pulsar in the J175229.06$-$284905.6 field is a sidelobe detection, as the timing position and our pointing center differ by about $8.5^{\prime}$. The refined position of PSR J1753$-$2851 also places it well within the error ellipse of the \textit{Fermi} source 4FGL J1753.2$-$2848, suggesting that it is very likely a $\gamma$-ray pulsar.

We also cross-checked whether our new pulsars show positional associations with \textit{Fermi}-LAT sources by cross-matching with the 4FGL catalog. None of the new discoveries lies within the $95\%$ localization radius ($r_{95}$) of a 4FGL source. All sources have separations $>7\arcmin$, except for PSR~J1720$-$2652, which lies $4.45\arcmin$ from 4FGL~J1720.6$-$2653c, slightly outside its $r_{95}=3.83\arcmin$. However, as indicated by the ``c'' suffix, this is a confused source located in a region where the diffuse background modeling is uncertain and multiple nearby sources make it difficult to disentangle the emission, thereby increasing the positional uncertainty. Therefore, the possibility of an association between PSR~J1720$-$2652 and 4FGL~J1720.6$-$2653c cannot be ruled out.

\section{Discussion and Conclusions} 
\label{sec:discussion}

We have conducted follow-up observations of a subset of pulsar candidates identified as circularly polarized sources in MeerKAT image-based searches of the Galactic bulge using the Ultra-Wideband Low (UWL) receiver of the Parkes radio telescope. Using candidates selected by \citet{frail_24} from the $\approx 173~\mathrm{deg}^2$ MeerKAT full-Stokes imaging survey of the Galactic bulge presented by \citet{cotton_25}, we followed up 16 sources and detected 9 pulsars.

Six pulsars are new discoveries including five MSPs (at least four of which are likely to be in binary systems). We also independently discovered PSR J1728$-$2608 and redetected two known pulsars. Among these three previously reported sources, two are binary MSPs, while the third has a relatively fast spin period of $\sim$85 ms. Extrapolating to all 30 candidates suggests that up to 16-17 MSPs might be discovered in total, which is equivalent to roughly one MSP per $\sim$10~deg$^{2}$. In fact, in a complementary work by Maan et al. (in prep.), five additional circularly polarized sources have been confirmed as MSPs, bringing the total to 13 MSPs so far. It is also interesting  to compare these results with MSP yield in untargeted large scale time-domain surveys such as The MPIfR-MeerKAT Galactic Plane Survey at L-band \citep[MMGPS-L;][]{padmanabh_23}, which covered about 936~deg$^{2}$ of the Galactic plane and discovered 20 MSPs i.e., one MSP per $\sim$47~deg$^{2}$. Moreover, MMGPS-L required nearly 800 hours of telescope time which corresponds to one MSP per 40 hours of integration. By contrast, the image-based approach used only $\sim$10 hours (corresponding to almost 55\% of the circularly polarized sources) of MeerKAT time to discover and rediscover 7 MSPs and 2 fast pulsars ($P<100 \rm \, ms$) which is only about 1.4 hr per MSP discovery. This difference in efficiencies is striking, and demonstrates that image-based searches, when combined with time-domain follow-up and searches, represent an exceptionally efficient use of scarce telescope resources.

\begin{figure}
    \includegraphics[width=0.99\columnwidth]{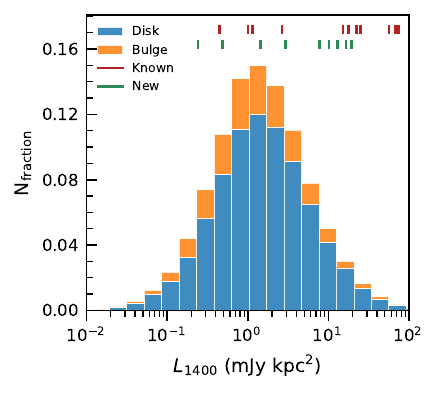}
     \caption{Simulated 1400-MHz luminosity ($L_{1400}$) distributions of MSPs from \texttt{PsrPopPy} population synthesis code. The stacked histograms in blue and orange show the Galactic disk (\(|b| < 1^\circ\)) and bulge (\(|\ell| < 4.5^\circ,\; 1^\circ < |b| < 20^\circ\)) MSP population normalized relative to the total disk and bulge population, respectively. Vertical markers indicate the luminosities of known MSPs detected in \citet{frail_24} in red and the luminosities of pulsars reported in this work (Table \ref{tab:msps}) in green.}
    \label{fig:luminosity}
\end{figure}

Several population synthesis studies predict that thousands of MSPs should exist in the bulge, possibly accounting for the Galactic-center GeV excess. Population study of MSPs by \citet{gronthier_18} predicts $\sim10^4$ MSPs with $L_{1400} \gtrsim 0.1$\,mJy\,kpc$^2$ should be concentrated in the bulge. \citet{ploeg_20} and \citet{berteaud_21} modeled an X-shaped bulge distribution with similar numbers required to produce the gamma-ray profile. Although all pulsars reported in this paper have DMs between $18$ and $329\,{\rm pc\,cm^{-3}}$, implying distances of only a few kiloparsecs based on the DM-derived distances. Therefore, these systems appear to lie in the foreground of the Galactic bulge, with a possible  exception of PSR~J1751$-$2943, which has the highest DM of $329\,{\rm pc\,cm^{-3}}$ among the pulsars reported here.  Only three MSPs within $\pm15^\circ$ of the Galactic center have ${\rm DM}>300\,{\rm pc\,cm^{-3}}$. One of these lies in the globular cluster Terzan~6 \citep{gao_24}. However, both DM distance models suggest $d \sim 4.9$\,kpc for PSR~J1751$-$2943, which is somewhat too short for it to be located in the Galactic bulge, estimated to lie at $\sim6.2$\,kpc along this line of sight. However, given the uncertainty in DM distance modeling, there is a possibility that this pulsar might belong to the population of the Galactic bulge. A promising test to confirm this would be to obtain reliable spin-down, $\dot{P}$  measurements. Since the spin-down rate is directly related to spin-down power $\dot{E}=4\pi^2 I \dot{P}/P^3$, $\gamma$-ray luminosity  and flux at $d\approx5$--$6$\,kpc can be constrained. This inferred $\dot{E}$ will then allow us to assess whether PSR~J1751$-$2943 is energetically consistent with MSP interpretation of the  $\gamma$-ray excess in the bulge. The lack of confirmed bulge MSPs in our sample suggests one or both of two possibilities---either the actual bulge population is smaller than these models predict, or the survey  is insufficiently sensitive to bulge MSPs under realistic propagation conditions.

\begin{figure}
    \includegraphics[width=0.99\columnwidth]{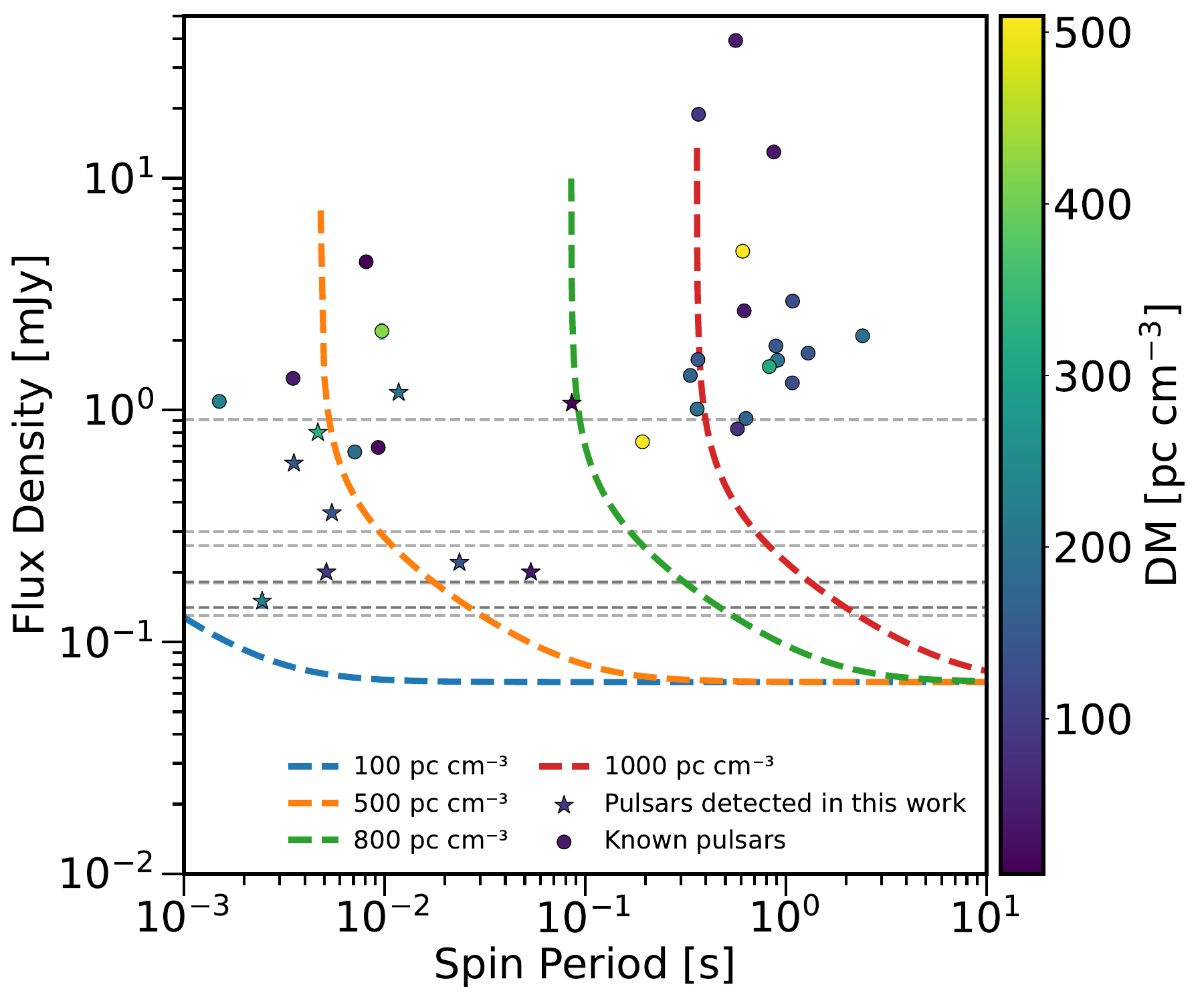}
     \caption{Sensitivity limits of the Parkes as a function of pulsar spin period for different DMs. The dashed curves are the sensitivity curves which show the minimum detectable flux density for DMs of 100, 500, 800, and 1000 $\rm pc \, cm^{-3}$ at  central frequency of 1250,  500MHz effective bandwidth, and a 1\,hour integration time. The effective pulse width in these calculations is $W_{\mathrm{eff}} = \sqrt{W_{\mathrm{int}}^2 + \tau_{\mathrm{scat}}^2}$, where $W_{\mathrm{int}}$ corresponds to a 5\% duty cycle and $\tau_{\mathrm{scat}}$ is the scattering timescale estimated using the empirical relation of \citet{bhat_04}. Overplotted are the pulsars detected in this work (stars) and previously known pulsars (circles) detected in \cite{frail_24}, with marker colors indicating their DM. Flux densities are taken from \citet{frail_24}, and only those previously known pulsars with measured fractional circular polarization in that study are included. The dashed horizontal lines indicate the flux densities of unconfirmed candidates.}
    \label{fig:senstivity_curves}
\end{figure}

To assess whether the MeerKAT image-based searches had the sensitivity to detect Galactic bulge MSPs, we conducted a population synthesis study using \textsc{PsrPopPy}\footnote{\url{http://samb8s.github.com/PsrPopPy/}} and simulated two MSP populations. One population was a disk population with $|b|<1^\circ$, and the other was a bulge population ($|l|<4.5^\circ$ and $1^\circ<|b|<20^\circ$). Figure \ref{fig:luminosity} shows the resulting distributions of $L_{1400}$. The observed $L_{1400}$ values of known MSPs detected in \citet{frail_24} and confirmed pulsars reported in this work fall mainly on the high-luminosity side of the simulated distributions ($1$--$70~\mathrm{mJy~kpc^2}$). This pattern is consistent with selection effects that preferentially reveal intrinsically brighter MSPs. Although the simulated bulge distribution does not appear markedly different from that of the disk, the presence of several mid-to-high $L_{1400}$ sources is more naturally explained by a disk+bulge scenario, as a disk-only population tends to underproduce such bright detections along these sightlines.

\begin{figure}
    \includegraphics[width=0.99\columnwidth]{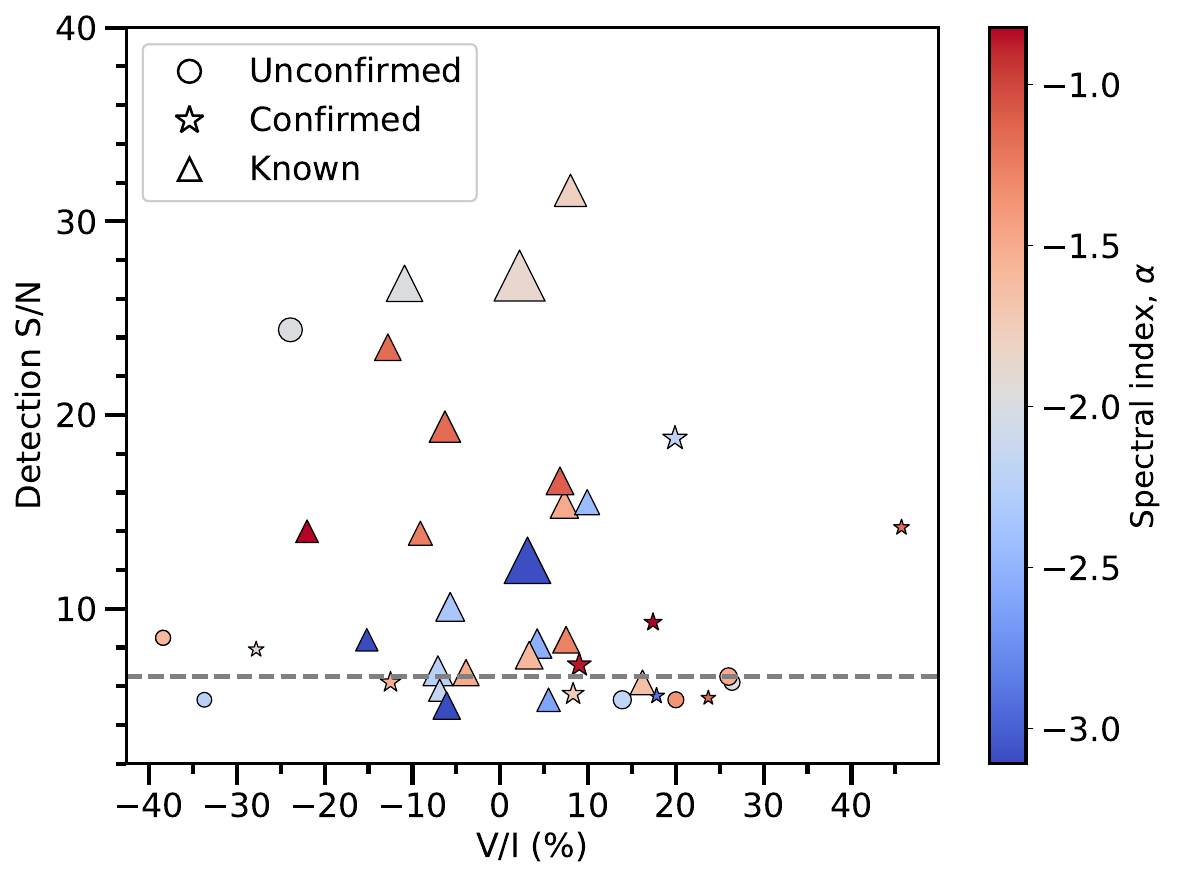}
     \caption{Fractional circular polarization ($V/I$) vs detection signal-to-noise ratio (S/N) for unconfirmed pulsar candidates in circles, newly confirmed pulsars in stars, and previously known pulsars in triangles. The marker size is proportional to the measured total intensity flux density ($I$), while the color scale represents the spectral index ($\alpha$). The horizontal dashed line corresponds to a low detection S/N of 6.0. All parameter values are taken from \citet{frail_24}.}
    \label{fig:cp_sn}
\end{figure}

The high efficiency of our search highlights the importance of combining full-Stokes interferometric imaging with polarization-based pulsar candidate selection and wide-band time-domain searches. In particular, circular polarization has been proven to be a strong discriminator for identifying promising pulsar candidates. Apart from this, the use of the UWL receiver has played a crucial role in discovering these pulsars. Its coverage from 704--4032\,MHz allows simultaneous access to both lower frequencies (700--1500\,MHz), where steep-spectrum pulsars are brighter (e.g., PSRs~J1811$-$3206 and J1818$-$3254 in our sample), and higher frequencies, where interstellar scattering is less severe (e.g. PSRs~J1723$-$2820 and J1751$-$2943). Four out of six new pulsars (PSRs J1720$-$2652, J1723$-$2820, J1751$-$2943, J1753$-$3331) reported in this work would likely have been undetectable with narrower L-band systems (1000-1500 MHz), as their pulse profiles are either strongly scattered below 1.5\,GHz or very faint, but are clearly detectable in wide-frequency band from 1000-2000 MHz or higher. Also, they are faint, steep spectrum sources, which also explains why these pulsars remained undetected in previous surveys. On cross-checking the pulsars' locations, we found that four of nine pulsars lie within the sky coverage of the HTRU-S LowLat survey ($|b| < 3.5^\circ$). Among these, PSRs~J1744$-$2424 and J1752$-$2849 were already detected in HTRU-S, but in the absence of timing their positions were uncertain within $\sim7$\,arcmin. We also found that PSR~J1753$-$3331 is present in the HTRU data, but with significant scattering and a low ${\rm S/N}=9.4$, which left it among the thousands of false positives of HTRU-S LowLat. Thanks to the wide-band capability of the UWL receiver, this pulsar was redetected and confirmed with higher ${\rm S/N}$ of 14. For PSR J1751$-$2943, the closest observation in HTRU-S is about 18 arc-minutes away, so we did not find this pulsar. These four pulsars are also present in the sky region of the Parkes multi-beam survey (PMPS), however given the shorter integration time (36 minutes), limited bandwidth and coarser frequency and time resolution, none of these pulsars are detectable in the PMPS data. The remaining four pulsars are either present in the regions covered by the PMPS or HTRU-S mid-latitude survey \citep[$|b|<15^{\circ}$][]{keith_10}, but given its short integration time of 8 minutes these pulsars remained undetected.

While we know why these pulsars were not detected in previous surveys, these new discoveries pose two additional questions: why have we only detected fast pulsars, and why the remaining 7 circularly polarized sources have not been confirmed as pulsars? To answer the first question, the high fraction of MSP discoveries in this work can be explained by a combination of selection and propagation effects. Slow pulsars are often easier to identify in L-band surveys toward the Galactic bulge, since scattering-induced pulse broadening is less likely to wash out their longer-period signals. In contrast, fast pulsars can remain hidden due to severe scattering below $\sim 1.5\,\mathrm{GHz}$.
The use of image-based searches along with the use of broad frequency coverage of the UWL receiver, extending to higher frequencies where scattering is reduced, allowed us to detect these MSPs. In order to explain why the remaining 7 circularly polarized sources fail to yield any pulsations, we compared the unconfirmed circularly polarized sources with sensitivity curves for Parkes at L-band, overlaid with both the newly discovered and previously known pulsars whose $|V|/I$ was reported in \citet{frail_24}. 

From Figure~\ref{fig:senstivity_curves}, it is evident that the flux densities of these pulsars lie in the range where other sources have been confirmed (0.15--0.9\,mJy). If the unconfirmed sources are indeed pulsars, one possibility for their non-detection is that they are intrinsically faint, highly scattered, and/or steep-spectrum objects. However, these flux densities are derived from imaging, and from PSRCAT we find that the calibrated pulsed flux densities of known pulsars in our sample are typically a factor of $\sim1.3$ lower than the imaging values. This implies that several of the unconfirmed sources may in reality have pulsed fluxes close to or below our L-band sensitivity once scattering is included. Therefore, $\sim1$\,hr Parkes observations may not be sufficient to detect them. They might be similar to PSRs J1720$-$2652 and J1753$-$3331, which are among the faintest detections presented in this work. On the other hand, 5 out of 7 unconfirmed sources have imaging $S/N < 6.5$ (see Figure \ref{fig:cp_sn}), supporting the assumption that they are faint, steep-spectrum, and highly scattered sources not visible at higher frequencies. Longer Parkes integrations ($>2$--$3$\,hrs) or short S-band (2--3\,GHz) observations ($\sim20$\,min) with MeerKAT would therefore be more promising for confirming these candidates.

Another reason could be that some/all of the sources in which pulsations were not discovered are sources of confusion. In a circularly polarized search, the primary sources of confusion are stars \citep{pritchard_21}, which can show polarization fractions varying from few tens to 100\%. Initial filtering for radio stars by \citet{frail_24} relied on classified stars (using \textit{Simbad}) and hence there is a possibility that fainter (and more distant) stars could be missed, which can manifest as confusion. Hence, we exploited the deeper Galactic plane observations, using the Dark Energy Camera Plane Survey \citep[DECam/DECaPS;][]{decaps} at optical wavelengths and the V\'{i}a L\'{a}ctea Survey \citep[VVV;][]{vvv} at near infrared (NIR) wavelengths (but more broadly, all publicly available data at these facilities). In the case of J1851$-$3456, we find a clear counterpart at the position of the radio source, indicating that it might be a possible star. For J1753$-$2930, we find a faint cataloged source with an optical $r$-band apparent magnitude of $r_{\rm AB} \approx 21.8$ mag (close to the noise floor in the AB magnitude system)  in the DECaPS data release 2, 0.28\arcsec\ offset, but within the 1-$\sigma$ uncertainty of the radio position. Given the faintness, deeper optical observations will be required to confirm this association. For the rest of the six sources in which pulsations were not detected, we do not find strong evidence for a counterpart within the 1-$\sigma$ region of the radio position.

In summary, this work provides compelling empirical evidence that circular polarization-based searches in the image plane, when combined with time-domain searches using wideband receivers, offer an effective alternative approach for discovering MSPs. With the advent of next-generation facilities such as the SKA \citep{keane_11}, DSA-2000 \citep{hallinan_19}, and the FAST Core Array \citep{jiang_24}, and the integration of deep, high-resolution imaging with optimized frequency strategies will be crucial for uncovering the true MSP population of the Galactic bulge, as well as expanding the known pulsar population more broadly.

\begin{acknowledgments}

We thank the anonymous referee for constructive comments that improved the manuscript. R.S.\ acknowledges the continuing valuable support of the Max Planck Society and support from NSF grant AST-1816904. AA and DLK were supported by NSF grants AST-1816492 and AST-1816904, and DLK is further supported by NSF grant AST-2511757. The MeerKAT telescope is operated by the South African Radio Astronomy Observatory, which is a facility of the National Research Foundation, an agency of the Department of Science and Innovation. The Parkes radio telescope (Murriyang) is part of the Australia Telescope National Facility, funded by the Australian Government for operation as a National Facility managed by CSIRO. We acknowledge the Wiradjuri People as the Traditional Owners of the Parkes Observatory site. The Australia Telescope Compact Array is also part of the Australia Telescope National Facility (\url{https://ror.org/05qajvd42}), and we acknowledge the Gomeroi People as the Traditional Owners of its site.
Basic research in radio astronomy at the U.S. Naval Research Laboratory is supported by 6.1 Base Funding.
\end{acknowledgments}

\vspace{5mm}
\facilities{MeerKAT, Parkes}

\software{astropy \citep{2013A&A...558A..33A,2018AJ....156..123A}, PSRCHIVE \citep{2004PASA...21..302H}, PEASOUP, RIPTIDE, PRESTO, Pulsar Survey Scraper \citep{2022ascl.soft10001K}, PyGEDM \citep{2021PASA...38...38P}.}

\bibliography{mkt_bulge}{}
\bibliographystyle{aasjournal}

\end{document}